\documentclass[%
 reprint,
 amsmath,amssymb,
 aps,
superscriptaddress,prl,floatfix]{revtex4-2}

\usepackage{graphicx}
\usepackage{dcolumn}
\usepackage{bm}

\usepackage{xcolor}
\usepackage{verbatim}

\begin{document}

\preprint{APS/123-QED}

\title{
Wall torque controls curvature-driven propulsion in bacterial baths
}

\author{Nicola Pellicciotta}
\affiliation{NANOTEC-CNR, Soft and Living Matter Laboratory, Piazzale A. Moro 5, Roma, 00185, Italy}
\affiliation{%
 Dipartimento di Fisica, Sapienza Università di Roma, Piazzale A. Moro 5, Roma, 00185, Italy}%
 
\author{Ojus Satish Bagal}%
\affiliation{%
 Dipartimento di Fisica, Sapienza Università di Roma, Piazzale A. Moro 5, Roma, 00185, Italy}%

\author{Maria Cristina Cannarsa}%
\affiliation{%
 Dipartimento di Fisica, Sapienza Università di Roma, Piazzale A. Moro 5, Roma, 00185, Italy}%

\author{Silvio Bianchi}%
\affiliation{NANOTEC-CNR, Soft and Living Matter Laboratory, Piazzale A. Moro 5, Roma, 00185, Italy}
\affiliation{%
 Dipartimento di Fisica, Sapienza Università di Roma, Piazzale A. Moro 5, Roma, 00185, Italy}%
\author{Roberto Di Leonardo}%
\affiliation{%
 Dipartimento di Fisica, Sapienza Università di Roma, Piazzale A. Moro 5, Roma, 00185, Italy}%
\affiliation{NANOTEC-CNR, Soft and Living Matter Laboratory, Piazzale A. Moro 5, Roma, 00185, Italy}

\date{\today}

\begin{abstract}
The persistent dynamics of active particles makes them explore extended portions of an obstacle's boundary during collisions. From impact to escape, the net applied forces depend on the curvature of the wall and increase in the presence of concave features.  
Here we systematically investigate the forces exerted by swimming bacteria on microfabricated structures, where the radii of curvature can be varied parametrically. We find that these micro-sails are propelled with a speed that scales linearly with curvature and is directed from concave to convex side along the axis of symmetry. By solving the collision problem for microswimmers with aligning torque interactions, we demonstrate that, unlike spherical active particles, curvature mainly affects cell orientation during sliding, leading to greater normal thrust on the concave side and an net applied thrust that scales linearly with curvature.
\end{abstract}

\maketitle

\textbf{\textit{Submitted to Physical Review Letters (PRL).}}

What is the mechanical action exerted by active particles on a suspended passive object is a problem that has attracted enormous interest both because of fundamental issues related to the generalization of the concept of pressure in non-equilibrium systems \cite{takatori2014swim,solon2015pressure,das2019local,yan2015force}, and because of obvious applications to the transport of colloidal cargoes\cite{koumakis2013targeted,palacci2013photoactivated,pellicciotta2023light,kaiser2014transport}.
The two issues are evidently interconnected, as the emergence of a net propulsion on a passive tracer necessitates a pressure distribution that, unlike in equilibrium conditions, is not homogeneous along the object's contour. This can be achieved by unbalancing the activity around the object \cite{pellicciotta2023colloidal} or, more often, by giving it an asymmetrical shape that results in the preferential accumulation of active particles in concave regions that act as propulsion hotspots \cite{angelani2010geometrically,di2010bacterial,sokolov2010swimming,vizsnyiczai2017light,pellicciotta2023light}. This latter geometric rectification is particularly efficient at high densities where jammed pockets of particles can cooperatively participate to propulsion through polar alignment induced by anisotropic interactions, such as in elongated swimmers \cite{kaiser2012capture,kaiser2014transport,di2010bacterial}. However, inter-particle interactions are not a necessary ingredient; asymmetric shapes can still lead to unbalanced pressure in the regime of non-interacting active particles. In this dilute regime, collisions of individual particles lead to the exploration of extended portions of the boundary with an overall mechanical action that depends on global characteristics of the wall contour, such as curvature. 
This has been extensively studied both theoretically and numerically for active Brownian particles, \textit{i.e.} self-propelled spherical objects with diffusive rotational dynamics that are typically unaffected by interactions with walls.
When confined by flat solid walls, these particles exert a mechanical pressure that can be derived from bulk properties by means of an equation of state, as for fluids in thermal equilibrium \cite{takatori2014swim,solon2015pressure,das2019local}
In the presence of curved boundaries, the equation of state breaks down and the pressure may exhibit a strong dependence on curvature\cite{fily2014dynamics,smallenburg2015swim,yan2015force,nikola2016active,mallory2014curvature,yan2018curved}. Numerical simulations demonstrate that even minimally curved objects can display directed motion due to a difference in active pressure between their concave and convex sides, resulting in net propulsion that grows linearly with the curvature\cite{mallory2014curvature,duzgun2018active}. However, it remains uncertain how relevant the predictions made with spherical particles and ignoring the effects of alignment on the walls are for real systems.
Strictly speaking active isotropic particles do not exist in nature. The propulsion mechanism inherently depends on the anisotropy of the particle's local environment, which is inevitably perturbed by the presence of obstacles. This results in reorientation effects at walls, even in the case of active particles of spherical shape \cite{ das2015boundaries, simmchen2016topographical, campbell2019experimental, rashidi2020influence, fins2024steer}.
For swimming microorganisms, reorientation at the wall is even stronger due to anisotropic shape and flagellar dynamics \cite{bianchi2017holographic,lauga2009hydrodynamics,bechinger2016active}. Accounting for torque interactions with walls is then essential to understand and predict rectification phenomena in experimentally relevant situations.\\
Here we study the mechanical action of swimming bacteria on microfabricated objects with a precise and systematically varied radius of curvature. Despite the strong alignment effects on the walls, we recover the ABP prediction of a linear dependence on the curvature of the net force, albeit for completely different reasons. 
We explain this trend by developing a model based solely on steric interactions between the bacteria and the walls, allowing us to quantify the forward impulse exerted by the bacteria on the micro-sails from their observed trajectories. This model effectively captures and reproduces key experimental trends, emphasizing the contrast with ideal active systems.

\begin{figure*}[t]
\centering
\includegraphics[width=0.95\textwidth]{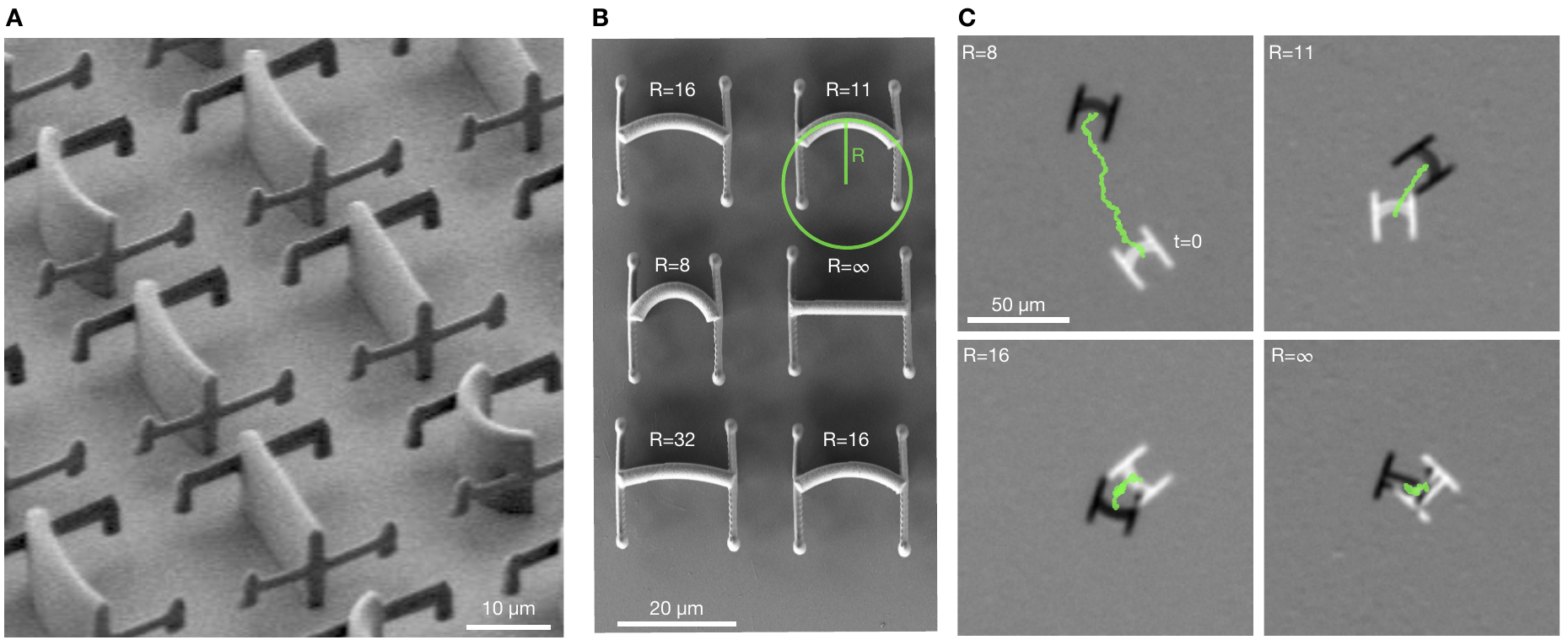}
\caption{ \textbf{Design and dynamics of micro-sails in a bacterial bath} 
(A-B) SEM images of SU8 micro-sails with varying curvatures: (A) side view and (B) top view. (C) Post-processed bright-field images showing the initial (white) and final (black) positions of the micro-sails, with the trajectory highlighted in green.}
\label{fig:fig1}
\end{figure*}

\textit{Results}-We used two-photon polymerization to fabricate microstructures from SU-8 photoresist \cite{maruo2008recent,vizsnyiczai2017light} that have the shape of a systematically curved wall. A SEM image of the fabricated structures is shown in Fig.~\ref{fig:fig1}A,B. These micro-sails have a fixed arch length of 20\;µm, height of 12\;µm, wall thickness of 2\;µm, and five different radii of curvature R = [$\infty$, 32, 16, 11, 8]µm. The micro-sails are kept vertical by H-shaped supports at the two extremes. The experiment is performed by adding the micro-sails and 5µL of \textit{E.coli} bacteria (refresh culture at OD=0.7) to an open chamber with a glass bottom filled with 1 mL of water and Tween 20 (0.02\%). The large volume and the high density of the micro-sails keep them fully sedimented on the bottom coverglass but without sticking. For all the experiments we used an \textit{E.coli} smooth-swimming strain lacking the tumbling mechanism. Consequently, cell reorientation arises solely from rotational diffusion or collisions with boundaries. Further details on the bacteria strain, micro-sail fabrication and sample preparation are provided in the Methods\cite{supmat}. The density of bacteria is kept low to avoid polar ordering and trapping of bacteria inside the curved regions of the micro-sail. The mean speed of the bacteria in the sample is $v_b=$30 µm/s, measured with Differential Dynamics Microscopy (DDM) from bright-field recording of the bacteria dynamics with 10X objective \cite{martinez2012differential,pellicciotta2023colloidal}.\\
The micro-sail dynamics are recorded for 7 minutes at 2 frames per second (fps) in bright field on an inverted microscope equipped with a 4X objective. We can extract their positions and orientations for several fields of view using a custom tracking algorithm based on template matching \cite{pellicciotta2023light}. In Fig.~\ref{fig:fig1}C, we show an example of the trajectories for a few micro-sails with different curvature radius. In Supplemental Material we provide a video of the acquired dynamics and tracking (Video1), and the dynamics imaged at higher magnification (Video2)\cite{supmat}.
\begin{figure*}[ht!]
\centering
\includegraphics[width=1.\textwidth]{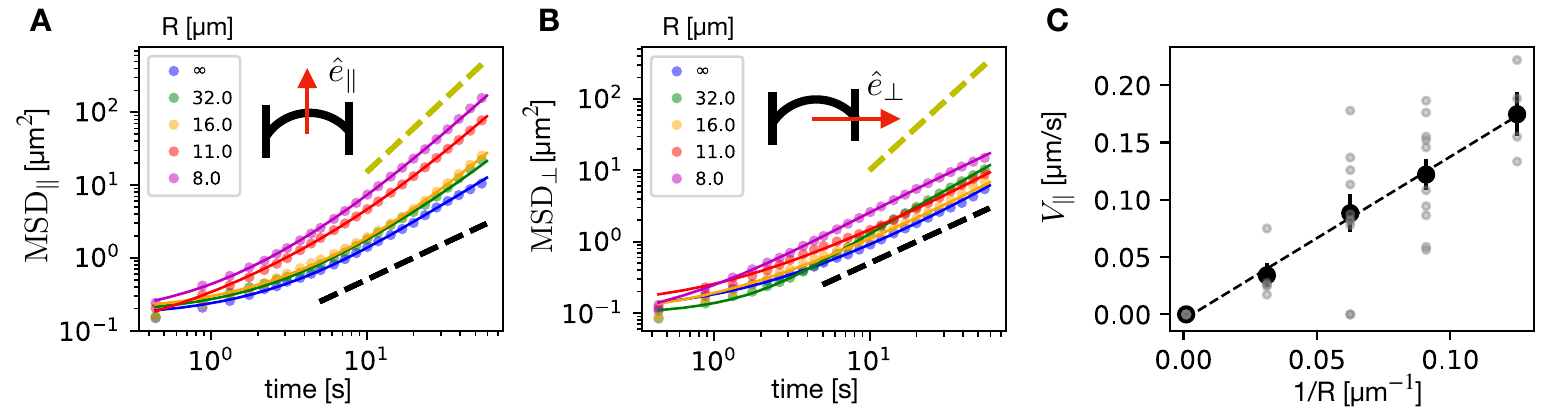}
\caption{ \textbf{Mean-Square Displacement reveals micro-sail propulsion.}
(A-B) The MSD along the main axis and perpendicular direction for five micro-sails with varying curvatures is well fitted by the model in Eq.~\ref{eq:eq2}. The black and yellow dashed lines indicate the diffusive and ballistic regimes, respectively. (C) The propulsion velocity increases linearly with the curvature of the micro-sails. Each gray dot represents data from a single micro-sail trajectory (30 in total), while the black dots indicate the mean values with standard error bars.}
\label{fig:fig2}
\end{figure*}
The propulsion of the curved micro-sails can be quantified by the net forward displacement $s_{\parallel}(t)= \int_0^{t} \boldsymbol{v}(t') \cdot \boldsymbol{\hat{e}_{\parallel}}(t') dt'$  where $\boldsymbol{v}(t')$ the instantaneous velocity at time $t'$ and $\boldsymbol{\hat{e}_{\parallel}}(t)$ is the main axis of symmetry of the structure pointing from the convex to the concave side of the wall. 
The longitudinal Mean Square Displacement is then calculated as MSD$_{\parallel}$=$\langle \Delta s_{\parallel}^{2} (t) \rangle$ for the 5 different curvatures and reported in Fig.~\ref{fig:fig2}A.
For the flat micro-sail (radius R= $\infty$), the longitudinal dynamics is diffusive at long times showing no net propulsion. However, as the curvature $k=1/R$ is progressively increased, the micro-sails experience a ballistic behaviour at long times, underlining a constant motion that increases as the micro-sail becomes increasingly curved. 
Due to the left-right symmetry of the micro-sail, we expect that the transverse force applied by bacteria will average to zero. Indeed, the micro-sails move diffusively along the perpendicular direction $\boldsymbol{\hat{e}_{\perp}}$ independently of the curvature, as shown by the transverse mean square displacement MSD$_{\perp} = \langle \Delta s_{\perp}^{2} (t) \rangle$ in Fig.~\ref{fig:fig2}B. 
The dynamics of the curved micro-sail can be modeled by a Langevin dynamics that includes both thermal fluctuations and interactions with swimming bacteria\cite{maggi2014generalized}:
\begin{align}
\boldsymbol{\dot{r}} &= \eta_{\parallel}\boldsymbol{\hat{e}_{\parallel}} +  \eta_{\perp}\boldsymbol{\hat{e}_{\perp}}, \nonumber\\
\dot{\phi} &= \eta_{z},
\label{eq:langevin}
\end{align}
where $\boldsymbol{r}(t) = (x(t),y(t))$ is the micro-sail position, $\phi$ is the orientation of the micro-sail around the z-axis,
and $\eta_{\parallel}$, $\eta_{\perp}$, $\eta_{z}$ are respectively the longitudinal, transverse and rotational noise. We assume these three noise components to be independent, \textit{i.e.}, $\langle \eta_{\alpha}(t)\eta_{\beta}(t')\rangle_{\alpha \neq \beta} = 0$, where $\alpha$ and $\beta$ represent one of the three symbols ($\parallel$,$\perp$,$z$). Each noise has two  independent contributions: $\eta_{\alpha}= \eta^T_{\alpha}+ \eta^A_{\alpha}$, with $\eta^T_{\alpha}$ representing the thermal noise and $\eta^A_{\alpha}$ describing the active component due to interactions with bacteria. The thermal noise has a zero average $\langle \eta^{T}_{\alpha} \rangle =0 $ and correlations $\langle \delta\eta^{T}_{\alpha}(t)\delta\eta_{\alpha}^{T}(t') \rangle = 2D^{T}_{\alpha}\delta(t-t')$, where $D_{\alpha}^T$ is the thermal diffusion coefficient and $\delta\eta_{\alpha}^{T} = \eta_{\alpha}^T - \langle \eta^{T}_{\alpha} \rangle$. 
Conversely, the active noise may have a non vanishing mean value $\langle \eta^{A}_{\alpha}(t) \rangle = V_{\alpha}(k)t$, depending on the sail curvature $k=1/R$.
Active fluctuations decorrelate exponentially with time $\langle \delta\eta^{A}_{\alpha}(t)\delta\eta_{\alpha}^{A}(t') \rangle = D_{\alpha}^{A}\exp(-|t-t'|/\tau)/\tau$, with $D^{A}_{\alpha}$ representing the active diffusion coefficient
and  $\tau$ the characteristic time that, for bacteria, is typically order of seconds. 
From Eq.~\ref{eq:langevin}, we can find the following expression for the mean square displacement and mean square angular displacement\cite{maggi2014generalized}:
\begin{equation}
\mathrm{MSD}_{\alpha}= 2D_{T}t + 2D_{a}[t-\tau(1-e^{-t/\tau})] +\\ V^{2}_{\alpha}(k)t^{2}
\label{eq:eq2}
\end{equation}
The above expression provides an excellent fit for the experimental data, see Fig.~\ref{fig:fig2}A,B. 
Remarkably, in the longitudinal direction, the propulsion speed $V_{\parallel}$ was found to increase linearly with curvature, reaching a maximum value of $V_{\parallel} \approx 0.20$ µm/s for the micro-sail with the highest curvature, Fig.~\ref{fig:fig2}C. In contrast, the active noise parameters showed no clear dependence on curvature. As expected, the extracted transverse speeds $V_{\perp}$ are very small compared with those along the main axis and show no dependence on curvature, see Supplemental Material Fig.~2\cite{supmat}. Regarding the angular speed $V_z$, we found that it is nonzero and consistently counterclockwise (when viewed from above the substrate). However, it does not exhibit any clear dependence on the curvature (see Supplemental Material Fig.~2D,E\cite{supmat}). This net angular speed may be explained by the tendency of \textit{E.coli} bacteria to swim counterclockwise over a solid surface \cite{lauga2006swimming}. 

\begin{figure*}[t]e
\centering
\includegraphics[width=0.95\textwidth]{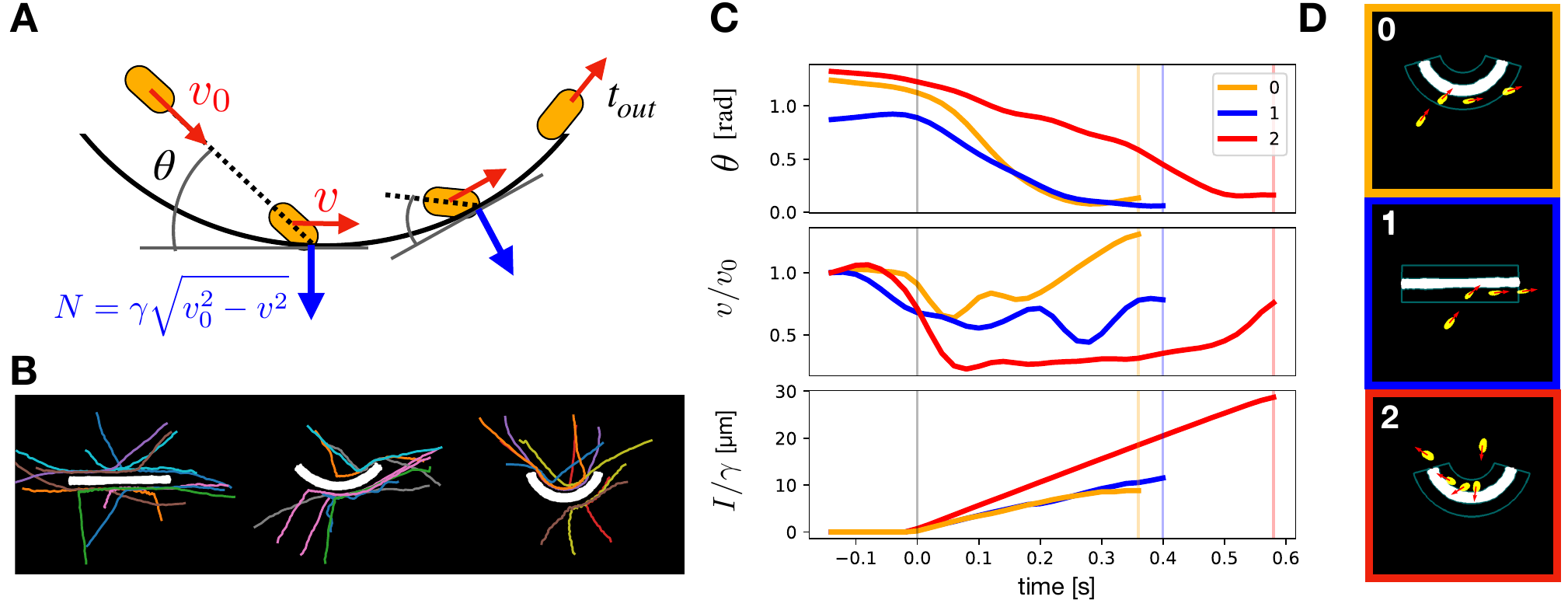}
\caption{ \textbf{ Measuring the force exerted by bacteria on micro-sails through trajectory analysis.}
(A) Schematic of a bacterium's trajectory along a curved surface.
(B) Experimental trajectories of bacteria interacting with fixed micro-sails recorded during a 30-second data acquisition.
(C) The angle $\theta$ between the direction of bacterial propulsion and the surface tangent, normalized velocity, and impulse ($I(t) = \int_0^{t} f(t') dt'$) for three bacteria colliding with micro-sails of negative, flat, and positive curvature. The vertical gray line marks the moment of collision, while the colored lines indicate when the bacteria detach from the micro-sail.
(D) Trajectories of the three bacteria, with their orientation indicated by red arrows}
\label{fig:fig3}
\end{figure*}

We will now introduce a minimal collision model that only considers self-propulsion, elongated shape and steric interactions between bacteria and walls. Following the derivation in the Supplemental Material Sec.II\cite{supmat}, we find that a cell swimming at speed $v_0$ away from the wall will slow down after collision and slide along the tangent of the wall with a speed $v$:
\begin{equation}
v = v_0\cos\theta
\end{equation}
where $\theta$ is the angle between the bacterial propulsion direction and the surface tangent, Fig\ref{fig:fig3}A.
Right after contact, the normal reaction from the wall will produce alignment with angular speed
\begin{equation}
\dot\theta=-\alpha\cos\theta\sin\theta
\label{eq:theta_flat}
\end{equation}
At each moment during sliding, the cell will exert a normal force on the wall given by
\begin{equation}
N=\gamma\sqrt{v_0^2-v^2}
\end{equation}
In this context, the normal force exerted by a single cell on a curved surface can be estimated solely from the kinematic quantities $v_0$ and $v$, which are readily obtained from cell tracking data.
This force can thus be found from the initial and instantaneous velocity without knowing the $\theta$ angle, which can be challenging to measure accurately in the vicinity of the micro-sail due to the wobbling of the cell.
For the whole duration of a collision event, the cell will transfer a forward impulse $I$ to the micro-sail given by
$I\simeq \int_0^{t_{out}} N(t) dt$, where $t=0$ and $t=t_{out}$ are the initial and final time of collision (residence time). We assumed that the curvatures are small enough that the normal force $N$ is always approximately aligned with the micro-sail axis $\hat{\mathbf e}_\parallel$.
At low Reynolds numbers, the impulse of a force translates to a displacement once it is divided by the viscous drag of the object receiving it.
Based on the discussion above, we can estimate the mean forward impulse transferred by a cell to a micro-sail during a collision event by tracking fluorescent bacteria colliding with micro-sails of variable curvature and fixed to the substrate glass.
Fig.~\ref{fig:fig3}B shows the tracked trajectory of many individual \textit{E.coli} cells during a single recording of 30s with a 60X objective, see Video3 in Supplemental Material\cite{supmat}.
As an illustrative example, Fig.~\ref{fig:fig3}C,D shows the dynamics of individual bacteria upon colliding with surfaces of three different radii of curvatures (R=-8,0,8 µm).
In the case of a surface with zero or negative curvature, the bacterium reorients itself to align with the surface (\(\theta \approx 0\)) and then either leaves the wall ($k<0$) or moves nearly parallel to it at its original speed, exerting minimal force (for a flat surface). Conversely, for surfaces with positive curvature, the bacteria take longer to align, and after that, resulting in a continuous application of force throughout their time on the surface, which leads to a greater total impulse being applied.
\begin{figure}[t]
\centering
\includegraphics[width=0.4\textwidth]{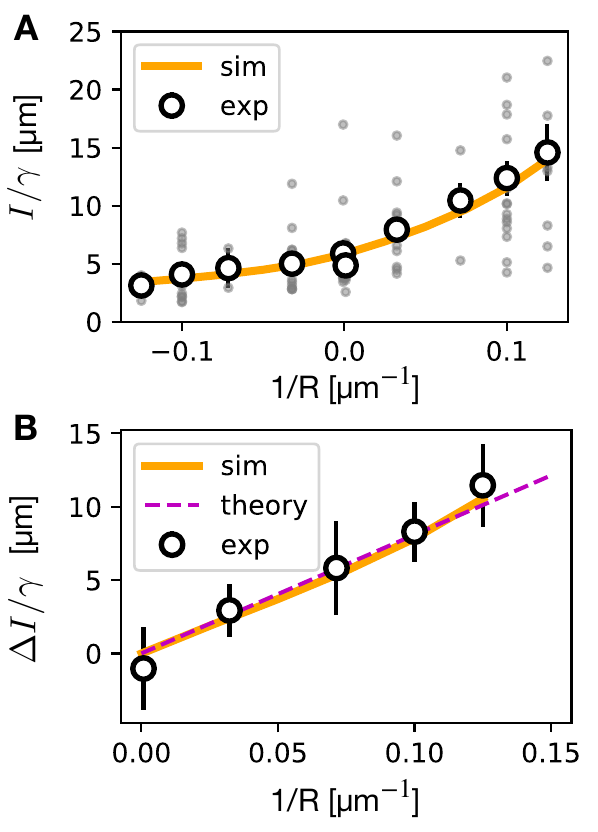}
\caption{ \textbf{ Net impulse on micro-sails increases linearly with curvature: experiments and model predictions.}
(A) Total impulse exerted by a single bacterium as a function of surface curvature. Gray circles represent data from individual bacterial trajectories, while white markers show the average. Error bars indicate standard error. (B) The net impulse on fixed micro-sails, $\Delta I(k) =I(k)-I(-k)$, increases linearly with curvature. Error bars represent propagation errors. Simulations are shown in orange for both panels. Dashed  magenta line represents the predicted net impulse from Eq.~\ref{eq:deltaI}. Both the simulation results and model predictions are scaled by a factor of 1.16 to best highlight the agreement of the trend with the experiments.}
\label{fig:fig4}
\end{figure}
From the set of bacterial trajectories, we computed the average impulse exerted by a bacterium for both positive and negative curvatures. As anticipated, the forward impulse increases with curvature, shifting from negative to positive, as depicted in Fig.~\ref{fig:fig4}A. Notably, the net forward impulse on a micro-sail, calculated as $\Delta I(k) = I(k) - I(-k)$, shows a linear increase with curvature, aligning with experimental observations of moving micro-sails, Fig.~\ref{fig:fig4}B. For the calculation of $\Delta I(k)$ , we assumed that the number of bacteria impacting the micro-sails do not depend on the curvature.
We then proceed to calculate the forward impulse on a micro-sail based on simulations of the angular dynamics of $\theta$, as described in Eq.~\ref{eq:theta_flat}. The effect of surface curvature is introduced as an additional term, leading to:
\begin{equation} 
\dot{\theta} = - \alpha \cos\theta \sin\theta + v_0 k \cos\theta 
\label{eq:theta_k} 
\end{equation}
Here, $v_0 \cos\theta$ represents the bacterium's velocity parallel to the surface, and $k = 1/R$ is the curvature. 
In the equation above, we excluded the effect of hydrodynamic trapping by convex walls \cite{sipos2015hydrodynamic}, as this effect diminishes significantly below a characteristic radius $R^*=50$µm. All micro-sails in this study have radii smaller than $R^*$. 
By direct Euler integration, we numerically calculated the dynamics of $\theta$ for bacteria colliding with a curved boundary of arc-length $l = 20$ µm. Starting angles are distributed with probability $P(\theta) \propto \sin(\theta)$  within the range $\theta_0 \in [\theta_{\rm{min}},\pi/2]$, where $\theta_{\rm{min}}$ depends on $\phi$, the initial angular position of the bacterium with respect to the osculating circle of the boundary, see Supplemental Material Fig.~3A\cite{supmat}. 
Each simulated trajectory stops when the bacterium leaves the surface (at time $t_{out}$), that happens in two situations: (i) when the cell reaches the end of the curved wall;
(ii) for negative curvatures, when the bacteria are aligned tangentially to the surface, \textit{i.e}. $\theta=0$. The simulated angular dynamics are illustrated in Supplemental Material Fig.~3B\cite{supmat}. We used  parameters obtained from experiments: $v_0 = 30$ µm/s, and  $\alpha$ extracted from experimental trajectories for the case of zero curvature ($R=\infty$). In this case, the equation Eq.~\ref{eq:theta_k} simplifies to $d\tan\theta/dt = -\alpha \tan\theta$, which predicts that the tangent of the angle $\theta$ decays exponentially to zero during reorientation ($\theta \rightarrow 0$), with a characteristic time $\tau_c = 1/\alpha$ \cite{bianchi2017holographic}. Experimental data confirm that $\tan\theta$ relaxes exponentially with an average characteristic time of $\tau_c = 0.12(0.03)$ s, see Supplemental Material Fig.~4\cite{supmat}. Then $\alpha = 1/\tau_c \approx 8$ s$^{-1}$. 
We then calculated the transmitted impulse applied on the boundary as $I/\gamma= \int_0^{t_{out}} v_0\sin\theta$,  for all hitting angles $\theta_0$ and initial angular position $\phi$ and averaged them. The resulting impulses are shown as a function of the curvature and superimposed to experimental data, Fig.~\ref{fig:fig4}A. 
We scaled these data by a factor 1.16 to best highlight the agreement of the trend with the experiments. The simulation predicts a net impulse $\Delta I(k) = I(k)-I(-k)$ that is also in good agreement with the experiments with fixed structures and is a linear function of the curvature Fig.~\ref{fig:fig4}B, thus validating our results with mobile structures again.\\ 
\textit{Discussion} -
The impulse generated by each bacterium can be analytically derived from Eq.~\ref{eq:theta_k} (Supplemental Material Sec. III\cite{supmat}):
\begin{equation}
\frac{I(k)}{\gamma v_0} = \frac{v_0 k}{\alpha} t + \frac{1}{2\alpha}\log \left[ \frac{1+\sin\theta_0}{1-\sin\theta_0}\frac{1-\sin\theta_1}{1+\sin\theta_1} \right]
\label{eq:eqI_theory}
\end{equation}
The second term represents the total impulse transferred by a cell to a flat surface during reorientation from initial angle \(\theta_0\) and final angle \(\theta_1\).
For positive curvatures, cells reorient to reach a final stationary angle  \(\bar\theta = \arcsin(v_0 k / \alpha)\) corresponding to the stationary solution (\(\dot{\theta} = 0\)) of Eq.~\ref{eq:theta_k}. The first term can be thus interpreted as the impulse transferred by a cell sliding at the stationary angle $\bar\theta$ for the entire residence time $t$. For negative curvatures cells reorient to reach $\theta=0$ and then leave the surface. The first term has no clear physical interpretation in this case and accounts for a faster reorientation dynamics due to the negative second term in Eq.~\ref{eq:theta_k}.
The total net impulse on the micro-sail, defined as \(\Delta I = I(k) - I(-k)\) is given by:
\begin{equation}
\frac{\Delta I(k)}{\gamma v_0}  =  \frac{v_0 k}{\alpha} \left[t(k) + t(-k)-\frac{1}{\alpha}\right] +O(v_0 k/\alpha)^3
\label{eq:deltaI}
\end{equation}
where we have assumed $\theta_1=\bar\theta$ for $k>0$ and $\theta_1=0$ for $k<0$ and the $\theta_0$ dependence cancels out.
This expression predicts a linear relationship between propulsion and curvature, provided the sum of the residence times \([t(k) + t(-k)]\) is only weakly dependent on $k$. 
Predicting the total residence time of a bacterium in advance is challenging, as it depends on both the curvature and the total surface length of the micro-sail. To get a rough estimate of the total residence time we can split the collision event in two main stages, alignment and sliding and add up their characteristic timescales 
\(t = t^a + t^s\). The alignment time (\(t^a\)), depends on the curvature and can be estimated from the bacterium's dynamics when it is far from alignment (\(\theta \approx \pi/2\)). Under these conditions, we can solve Eq.~\ref{eq:theta_k} with $\sin \theta \approx 1$, and find that the angle decreases exponentially as: $\theta(t) \approx \pi/2 - (\pi/2 - \theta_0)\exp[(\alpha - v_0 k)t]$. From this, an expression for the alignment time is derived as $t^a \approx A/(\alpha - v_0 k)$, where \(A\) is a fitting constant.
The sliding time $t_s$ is the time spent moving along the surface after reorientation to a positive stationary angle $\theta_1$, which occurs exclusively for curvatures $k \geq 0$.    During this second stage the bacterium is nearly aligned with the surface and moves with a velocity \(v \approx v_0\). The time \(t^p\) can then be estimated as 
$t^p \propto l^p/v_0$ with the parameter \(l^p\) being an effective path length. Combining these contributions yields a residence time expression $t \approx A/(\alpha - v_0 k)+ l^p/v_0$ that aligns well with our simulations. The fitting parameters are found to be \(A \approx 2\) and \(l^p \approx 4\,\mu\mathrm{m}\) (see Supplemental Material Fig.~5\cite{supmat}). This formulation also predicts that the sum of residence times on both sides of the micro-sail remains nearly constant for small curvatures: $[t(k) + t(-k)] \approx 2A/\alpha + l^p/v_0 + O\left(v_0 k/\alpha\right)^2$. This near constancy well match with the observed linear relationship between net propulsion and curvature in our experiments. Figure~\ref{fig:fig4}b compares the predicted net impulse, computed using the fitted parameters, with the experimental data and simulation results, showing excellent agreement.\\
As opposed to ideal active particles undergoing free angular diffusion, we found that the presence of a wall torque results in transient interactions events during which anisotropic particles align to the wall and eventually always escape. 
In particular, reorientation occurs faster on the convex side, causing the cell to align completely and leave the surface, while on the concave side, it proceeds more slowly to a finite pitch angle, keeping the cell attached and pushing on the surface until it reaches the wall's end.
Despite this complex behavior, we predict that the overall propulsion exerted by bacteria on the two sides of a curved wall is proportional to its curvature which is confirmed by both experiments and simulations. These findings extend prior theoretical analyses of active pressure to encompass the more realistic scenario of torque interactions at boundaries. Moreover they have important implications for the practical design of passive structures with shapes optimized to fully exploit bacterial thrust, opening up applications in bio-hybrid microrobotics \cite{volpe2024roadmap}.

\textit{Acknowledgments}-The research leading to these results has received funding from the European Research Council under the ERC Grant Agreement No. 834615 (R.D.L.). O.S.B. and R.D.L have received funding from Marie Sklodowska-Curie actions under Grant agreement no. 812780. 


\bibliography{biblio}

\end{document}